\documentstyle[12pt]{article}

\title{ Super Star Products and  Quantum Superalgebras } 

\author{\bf {M. Mansour \thanks{
e-mail: fguedira@fsr.ac.ma}}\\\\ 
Laboratoire de Physique Theorique \\ 
Departement de Physique\\Universite of Mohamed V.\\B. P. 1014 Rabat\\ Morocco}

\newtheorem {Definition}{Definition}
\newtheorem {Proposition}{Proposition}
\begin{document}           

\maketitle

{\bf Abstract}: We prove that the super star product
on a  Poisson Lie supergroup leads to the structure of  
quantum superalgebra ( triangular Hopf superalgebra) on the super quantized 
enveloping algebra of the  Lie superalgebra of the Lie 
supergroup and that equivalents super star products generate isomorphic quantum
superalgebras. 

\vfill
\begin{flushright}
Typeset using \LaTeX
\end{flushright}

\break 

\section{Introduction} 

\hspace*{0.5cm}  The development of the quantum inverse scattering method (QISM) 
\cite{fad} intended for investigations of  integrable models of the 
quantum field theory and statistical physics  gives rise to some 
interesting algebraic constructions. These investigations allow to select 
a special class of Hopf algebras now known as quantum groups and quantum 
algebras \cite{dri,jim}.
The nice R-matrix formulation of the quantum group theory \cite{frt}, based 
on the fondamental relation of QISM (the FRT relation) has given an 
additional impulse to these investigations.
The extension of the activity on quantum groups to the field of supersymmetry was started with the paper of Manin \cite{mani1}, where the standard multiparametric quantum deformation of the supergroup $GL(m/n)$  was introduced. The study of the superalgebra in duality with the standard multiparametric deformation of $GL(m/n)$ was given in \cite{tah}. Quantum superalgebras 
appeared naturally when the quantum inverse scattering method  
was generalized to the super-systems \cite{ks1}. Related R-matrix were considered in \cite{kr,kt} and simple examples where 
presented in \cite{ku}.   The works \cite{fs,ckl,ck} are devoted to the 
q-bosonization of the q-superalgebras.The properties of 
  the quantum superalgebras $ U_{Q}(A(m,n)), U_{Q}(B(m,n)), U_{Q}(C(n+1))$ and $U_{Q}(D(m,n))$  when their deformation  parameter $Q$ goes to a
root of unity, are investigated  in \cite{ma11,ma21}. 
  Quantum supergroups, were investigated in \cite{ck,pu,pp}.\\
\hspace*{0.5cm}  As  is well known, quantum groups can be seen as noncommutative 
	 generalizations of  topological spaces which have a group structure. Such a structure induces an abelian Hopf algebra  structure \cite{abe} on the algebra of smooth functions on the group.
 Quantum groups are defined then as a non abelian Hopf algebras \cite{tak}
 . A way to generate them consists of deforming the abelian Hopf 
 algebra of functions into a non abelian one (*-product), 
 using the so called  deformation quantization or 
 star-quantization \cite{b1,b2,fls,mo1,mo2}.  \\
\hspace*{0.5cm}  A star-quantization method is used also to give an
 h-deformed algebra (quantum Lie algebra) in \cite{ma2} and to realize
  both q-deformed Virasoro and $su_{q}(2)$ algebras in \cite{am} and a
  deformed Yangian algebra in \cite{am2}. The notion of a super star-product
   on a symplectic flat supermanifold is investigated in \cite{jbk} and a
    deformation-quantization of Fedesov type os super Poisson bracket is
     given in \cite{bord}. The purpose of the present paper is to shows that
      the quantum supergroups can be generated  by deforming
 the graded (abelian) Hopf algebra of super functions structure into 
  a  non  graded-abelian one  ( super star-product). This quantization 
 technique gives a deformed product once  a Poisson 
 superbracket on the superalgebra of super smooth functions is given. 
 In order to ensure that the deformed superalgebra is a Hopf superalgebra,
  namely a quantum supergroup, the starting supergroup
$G$ has to be endowed with a super Lie-Poisson structure. Finally,   using the 
 duality procedure, this quantization leads to the structure of the  quantum 
 superalgebra on the super quantized enveloping algebra of the 
 Lie superalgebra corresponding to the above  Lie supergroup $G$ .\\
\hspace*{0.5cm} This paper is organized as follows; the second section is devoted to a review of basic definitions of  Lie bisuperalgebras and  Lie-Poisson supergroups. In the third section we show the main result
that states that a super star product on a Lie-Poisson supergroup leads to the structure of a quantum superalgebra  on the quantized enveloping algebra of the  Lie superalgebra corresponding to the above
supergroup,  we give a vector representation of the super quantum Yang-Baxter equation and  we show that  equivalents super star products generate  isomorphic quantum superalgebras. 
\section{\bf Basic Definitions} 
\hspace*{0.5cm}  Let us first recall some properties of the  vector superspaces
and Lie superalgebras on the complex number field.

 If 
${\large g}$ is a  vector super space then  ${\large g}= {\large g}_{0}  \oplus {\large g}_{1}$, where we refere to 
${\large g}_{0}$  and $ {\large g}_{1}$ as the even and odd subspaces of ${\large g}$ respectively. We define the operator index 
$$ \mid \mid : \,\,\,\,\,{\large g}\longrightarrow \{ 0,1 \}$$
for the homogeneous elements of ${\large g}$ by 
 $$ \mid x  \mid  = 0  \,\,\, if \,\,\, x \,\,\, \epsilon \,\,\,\,\, {\large g}_{0} $$
 $$ \mid x  \mid  = 1 \,\,\,if \,\,\,\,\,x \,\,\, \epsilon \,\,\,{\large g}_{1} $$
and call $(-1)^{\mid x  \mid }$ the parity of $ x $ . 
The dual 
${\large g}^{*}$ inherts a natural super gradation   ${\large g}^{*}= {\large g}^{*}_{0}  \oplus {\large g}^{*}_{1}$, with ${\large g}^{*}_{a}$ is isomorphic to the dual of ${\large g}_{a}$ $(a = 0, 1)$ . On the tensor product ${\large g} \otimes {\large 
g}$, there exists a natural 
super gradation induced from that of ${\large g}$, where the parity of $x \otimes y $ is related to those of the homogenous elements $ x, y  \in  {\large g} $ through 
$$ (-1)^{\mid x \otimes y \mid }= (-1)^{\mid x \mid + \mid y \mid }.$$ 
It is also useful to define the twisting map 
$$ T : {\large g} \otimes {\large g} \longrightarrow {\large g} \otimes {\large g}$$
by
 \begin{equation} 
T(x \otimes y )= (-1)^{\mid x \mid \mid y \mid }(y \otimes x )
\end{equation}
for all homogenous $x, y$  $ \in  \,\,\, {\large g}$ ; this definition is extended by linearity to all ${\large g} \otimes {\large g}$.\\
	A Lie superalgebra structure on ${\large g}$ is provided by a linear mapping $$ \lbrack,\rbrack : {\large g} \otimes {\large g}\longrightarrow   {\large g}$$
satisfying the requirement of super Jacobi identity and super antisymmetry. In order to express them it is 
useful to introduce a basis $ \{ X_{i} \}$ in ${\large g}$  and structure constants defined by
$$
\lbrack X_{i} , X_{j} \rbrack = C^{k}_{ij}  X_{k} .$$
Then the structure constants have to satisfy \\
$$  C^{k}_{ij} = 0~~ \mbox{whenever}~~ \mid X_{i}\mid + \mid X_{j} \mid \neq \mid X_{k} \mid (mod 2) $$
$$ C^{k}_{ij}= (-1)^{\mid i \mid \mid j \mid} C^{k}_{ji}~~ \mbox{(super antisymmetry)} $$
$$(-1)^{\mid i \mid \mid l \mid} C^{k}_{ij} C^{m}_{kl}+(-1)^{\mid i \mid \mid j \mid} C^{k}_{jl} C^{m}_{ki}+(-1)^{\mid j \mid \mid l \mid} C^{k}_{li} C^{m}_{kj} = 0  ~~\mbox{(super Jacobi identity )}$$.\\
A Lie bisuperalgebra structure on ${\large g}$ is given by a linear mapping $$ \phi : {\large g}  \longrightarrow   {\large g} \otimes  {\large g}$$
  $$ \phi( X_{i})= f^{kl}_{i}X_{k} \otimes X_{l}$$
where $ \phi $ has to satisfy several  requirements. First of all it makes the dual linear space $ {\large g}^{*}$ a Lie superalgebra i.e :
 $$ f^{ij}_{k} = 0 $$ 
whenever $ \mid X_{i} \mid + \mid  X_{j} \mid  \neq  
 \mid X_{k} \mid  (mod 2) $ and

$$(-1)^{ \mid k \mid \mid m \mid } f^{kj}_{i} f^{lm}_{j}+(-1)^{ \mid l \mid \mid k \mid} f^{lj}_{i} f^{mk}_{j}+(-1)^{ \mid m \mid \mid l \mid} f^{mj}_{i} f^{kl}_{j}=0. $$ 
 $ \phi $ must be a superalgebra 1-cocycle
$$ \phi \lbrack X, Y \rbrack = ad_{X} \phi(Y)- (-1)^{\mid X \mid \mid Y \mid} ad_{Y} \phi(X). $$ 
A cobondary Lie superalgebra is a pair $({\large g},r)$ where  ${\large g} $ is a Lie
 superalgebra and $ r $ \,\,\,$ \in \,\,\, (({\large g}_{0} \otimes {\large g}_{0})
 \oplus  ({\large g}_{1} \otimes {\large g}_{1}))$ such that for every  $ X_{i}\,\,\,  
\in \,\,\, {\large g} $ we have $$\phi( X_{i})= \lbrack r , 1 \otimes X_{i}+   X_{i} \otimes 1 \rbrack $$
where the even element $r$ satisfies the generalized classical Yang-Baxter equation 
\begin{equation}        
\lbrack [\![  r,r  ]\!] ,1 \otimes X_{i} \otimes 1+ X_{i} \otimes 1 \otimes 1 +  1\otimes 1 \otimes X_{i} \rbrack = 0 
\end{equation}

and the super Schouten bracket is defined as follows 
$$[\![  r,r  ]\!] = \lbrack  r_{12},r_{13}  \rbrack + \lbrack  r_{12},r_{23}  \rbrack + \lbrack  r_{13},r_{23}  \rbrack .
 $$
The coboundary superbialgebra with the r-matrix satisfying the modified classical Yang-baxter equation describe infinitesimaly Poisson-Lie supergoups which will be defined latter.
We now make the followimg definitions:
\begin{Definition}
A super quantized universal enveloping algebra is a topological Hopf superalgebra B with 
a bijective antipode over the ring of formal series $C\lbrack \lbrack  h \rbrack \rbrack$
, complete with respect to the h-adic topology and such that
 $\frac{\rm B}{\rm hB}$ is the universal enveloping algebra $U({\large g})$ of some Lie
 superalgebra $ {\large g} $.
\end{Definition}
Let $W= W_0 + W_1 $ be a differentielle supermanifold and $Fun(W)= Fun_{0} (W)\oplus Fun_{1} (W)$ be the algebra of supersmooth functions on $W$; $f \in Fun_{0} (W) [Fun_{1} (W)]$ is said to be homogenous of even [odd] parity.

\begin{Definition}
A super Poisson bracket $\{, \} $ on $Fun(W))$ is a bilinear operation assigning to every pair of functions $f,g \in Fun(W)$ a new function ${f,g} \in Fun(W)$, such that for homogenous functions satisfies  the following conditions:\\
i-Graded preserving
\begin{equation}\label{r1} deg (\{f, g \} = deg(f) + deg(g) \end{equation}
ii-Super skew-symmetry
\begin{equation} \label{r2}\{ f, g\} = - (-1)^{(deg(f))(deg(g))} \{ g, f \}\end{equation}
iii-Graded Leibniz rule
\begin{equation}\label{r3} \{f, gh \}= \{f, g\}h + (-1)^{deg(f) deg(g)} g \{f,h \} \end{equation}
iv-Super Jocobi identity
\begin{equation}\label{r4} (-1)^{deg(f) deg(h)} \{ f, \{g, h\}\} + (-1)^{deg(g) deg(f)} \{ g, \{h, f \} \} + (-1)^{deg(h) deg(g)} \{ h, \{f, g\}\}  =0 \end{equation}

\end{Definition}
Since the conditions [\ref{r1}-\ref{r4}] are just the axioms of superalgebras, the space 
$Fun(W)$ endowed with the super Poisson bracket becoms a Poisson superalgebra and $W$ a Poisson supermanifold.
\begin{Definition}
A Poisson Lie supergroup is a Lie supergroup G provided with a Poisson superbracket $ \{,\} $ such that the comultiplication 
$$\Delta: Fun(G) \longrightarrow Fun(G) \otimes Fun(G) $$
is a morphism of Poisson superbrackets:
$$\Delta \{\phi, \psi \} = \{ \Delta (\phi), \Delta(\psi)\} $$
\end{Definition}
 where the  Poisson superbracket on $Fun(G) \otimes Fun(G) $  is defined as 
\begin{equation}   
 \{ \phi_{1} \otimes \psi_{1}, \phi_{2} \otimes \psi_{2} \}=(-1)^{\mid \phi_{2} \mid \mid \psi_{1} \mid  }
  (\{ \phi_{1}, \phi_{2} \} \otimes \psi_{1} \psi_{2} + \phi_{1} \phi_{2} \otimes  \{ \psi_{1},\psi_{2} \}).
\end{equation}
and the following rule for the multiplication of graded tensor products should be used:
$$ (\psi_{1} \otimes \psi_{2}) (\phi_{1} \otimes \phi_{2}) = (-1)^{deg(\psi_{2}) deg(\phi_{1})}(\psi_{1}\phi_{1}\otimes \psi_{2}\phi_{2}) $$

\section{Super Star Products and  Quantum Superalgebras} 

\subsection{Triangular  Hopf Superalgebra Structure}

\hspace*{0.5cm}  Let $ G=  G_{0} \oplus G_{1} $ be a  Lie supergroup , ${\large g}$
 its Lie superalgebra. The enveloping algebra of the Lie superalgebra ${\large g}$ is
 defined \cite{kos} to be the tensor algebra
 $T({\large g})= \oplus_{k=0}^{\infty}{\large g}^{\otimes k }$, modulo the ideal I in $T({\large g})$  generated by all elements in $T({\large g})$ of the form 
\begin{equation}
 x \otimes y - (-1)^{\mid x \mid \mid y \mid  } y \otimes x
- \lbrack  x,y  \rbrack 
\end{equation}
for $ x, y  \in {\large g} $ \\
As in the classical case the Poincare-Birkoff-Witt theorem is valid for $ U({\large g})$,
 indeed for ${\large g}={\large g}_{0} \oplus {\large g}_{1}$ and if $ \{(e_{i}), \,\,
 i= 1,2,.....n \}$ is a basis of ${\large g}_{0}$ and $\{(v_{i}),\,\,
  i= 1,2,.....m \}$ is a basis of ${\large g}_{1}$ then a basis of $ U({\large g})$ is given by 
\begin{equation}  
      e_{1}^{k_{1}}.......e_{n}^{k_{n}}.v_{i_{1}}.....v_{i_{j}}
\end{equation}
where $k_{1},......k_{n} \in  N $ and $ 1 \le i_{1} \le ...,i_{j} \le m $. \\

Let 1 be the identity of the enveloping superalgebra. Then the morphism of degree zero ${\large g}$ into $ U({\large g}) \otimes U({\large g}) $ given by 

\begin{equation}        
x \longrightarrow  x \otimes 1 + 1 \otimes x
\end{equation}
extends to a morphism of degree zero 
\begin{equation}        
\Delta_{0}: U({\large g}) \longrightarrow  U({\large g}) \otimes U({\large g}).
\end{equation}
We note that for a  bisuperalgebra $ A= A_{0} \oplus A_{1}$, the coproduct preserves the parity; namely, one has 
$$ \Delta : A \longrightarrow A \otimes A  $$ 
$$ \Delta : A_{0} \longrightarrow A_{0} \otimes A_{0} +  A_{1} \otimes A_{1}$$ 
$$ \Delta : A_{1} \longrightarrow A_{0} \otimes A_{1} +  A_{1} \otimes A_{0}.$$ 
The antipode of the  enveloping superalgebra is defined as an homogenous   bijective map of degree zero  
\begin{equation}        
S_{0}: U({\large g}) \longrightarrow  U({\large g}) 
\end{equation}
such that for any $ x \in  \,\,{\large g}$ we have
\begin{equation}        
S_{0}(x)= -x
\end{equation}
and for $ u, v \in U({\large g})$ we have 
\begin{equation}        
S_{0}(u v)= (-1)^{\mid u \mid \mid v \mid  }S_{0}(v)S_{0}(u).
\end{equation}
Now let $ r  \,\, \in (({\large g}_{0} \otimes {\large g}_{0}) \oplus  ({\large g}_{1} \otimes {\large g}_{1}))$
be a solution of the super classical Yang-Baxter equation.
\begin{equation}
 [\![  r,r  ]\!] = 0 
\end{equation}
 Then the Lie  bisuperalgebra structure on {\large g} is given by the
superalgebra 1-cocycle
$$ \delta: {\large g} \longrightarrow  {\large g} \otimes {\large g}$$
\begin{equation}        
 x \longmapsto (ad_{x}\otimes 1 + 1 \otimes ad_{x})r
\end{equation}
where $ ad_{x} $ stands for the adjoint representation
and the super Poisson-Lie structure on Lie supergroup G is given by \cite{And}
\begin{equation}        
 \{ \phi , \psi \}= \sum_{i,j} \,\, (-1)^{\mid \phi \mid \mid j \mid}
r^{ij} ( X_{i}^{r}(\phi) X_{j}^{r}(\psi) - X_{i}^{l}(\phi) X_{j}^{l}(\psi))
\end{equation}
where $X_{i}^{r}= (R_{g})_{*}X_{i}$ and $X_{i}^{l}= (L_{g})_{*}X_{i}$ are 
the right and left vectors fields on the supergroup G, $(X_{i})$ is a 
basis of ${\large g}$  with  $(R_{g})_{*}$ and $(L_{g})_{*}$  the derivative 
maping corresponding to the right and left translation respectively .\\
If we denote by $R(G)( L(G))$ the set of all right(left)-invariant vector fields 
on $ G $, then using  elementary properties of derivative mapings \cite{dw} one may show
 that each of $L(G)$ and $R(G)$ is a  vector superspace with a bracket operation that 
satisfies the super Jacobi identity. Since every element of L(G) or R(G) \,\, is 
completely determined by its value at the identity element of G  it follows that
 $ L(G)$ and $ R(G)$ are isomorphic to the Lie  superalgebra (the tangent space to G at
 the identity (e)) .\\
Such morphisms can be extended  to  graded algebra morphisms 
\begin{equation}  
U({\large g}) \longrightarrow D^{l}(G) 
\end{equation} 
\begin{equation}        
 A \longmapsto A^{l}
\end{equation}
\begin{equation}   
U({\large g}) \longrightarrow  D^{r}(G) 
\end{equation}
\begin{equation}        
 A \longmapsto A^{r} 
\end{equation}
where $D^{l}(G)$ and $D^{r}(G)$ are respectively the superalgebra of left-invariant differential operators and the superalgebra of right-invariant differential operators, 
such that  the action of $U({\large g})$ on ${\bf F}(G)$ will be  given by 
\begin{equation}  < X, Y^{l}( \phi)  >= < X Y ,  \phi  > \end{equation}
\begin{equation}< X, Y^{r}( \phi)  >=(-1)^{\mid X \mid \mid Y \mid } <  S_{0}(Y)X  ,  \phi  >.  \end{equation}
We now make the following definitions \\
\begin{Definition}
A super star product on the  Poisson Lie supergroup is a bilinear map 
$${\Large F}(G) \times {\Large F}(G) \longrightarrow  {\Large F}(G) \lbrack \lbrack h \rbrack \rbrack  $$
\begin{equation}        
(\phi, \psi ) \longmapsto \phi * \psi = \sum_{j} h^{j} C_{j}(\phi, \psi )
\end{equation}
such that \\
i) when the above map is extended to ${\Large F}(G) \lbrack \lbrack h \rbrack \rbrack  $,  it is formally associative 
 \begin{equation}       
 (\phi * \psi)* \chi  = \phi *( \psi* \chi)
\end{equation}
ii)  $C_{0}(\phi, \psi ) = \phi. \psi = (-1)^{\mid \phi \mid \mid \psi \mid} \psi. \phi$\\
iii) $C_{1}(\phi, \psi )= \{ \phi, \psi \}$\\
iv) the two-cochains $C_{k}(\phi, \psi )$ are bidifferential operators , homogeneous of degree zero on ${\Large F}(G)$ .
\end{Definition}

The problem is  to get a  super star-product on the super group $G$ such that the compatibility relation 
\begin{equation}        
 \Delta (\phi * \psi)  = ( \Delta ( \phi) * \Delta ( \psi))
\end{equation}
is satisfied.
 The super star-product on the right side is canonically defined on ${\Large F}(G) \otimes {\Large F}(G) $ by 
\begin{equation}        
 (\phi \otimes \psi)* (\phi^{'} \otimes \psi^{'} )  =  (-1)^{\mid \psi \mid \mid \phi^{'} \mid} 
(\phi * \phi^{'})\otimes (\psi * \psi^{'} ).
\end{equation}
{\bf Remark}:
If all $C_{k}$ are a left (right)-invariant even bidifferential  operators 
then the corresponding super star product is called  left (right)-invariant .
\begin{Definition}
Two super star-products $ *_{1} $ and $ *_{2}$ defined on the supergroup $G$ are said to be formally equivalent if there exists a series
\begin{equation}        
T = id + \sum_{i=1}^{\infty} h^{i}T_{i}
\end{equation}
where the $T_{i}$ are  even differential operators , such that 
 \begin{equation}       
 T(\phi *_{1} \psi)  = T(\phi) *_{2} T( \psi). 
\end{equation}

\end{Definition}
Thanks to the  morphisms(18,20) , we see that if $C_{i}$ is a left-invariant even two cochain then there is an homogeneous element of degree zero $F_{i} \in \,\,\, U({\large g}) \otimes U({\large g}) $ such that: 
\begin{equation}        
 C_{i}^{l}(\phi ,  \psi)= F_{i}^{l}(\phi \otimes  \psi).
\end{equation}
Similarly for the right invariant even two cochain there exist an homogeneous element of degree zero $ H_{i} \in U({\large g}) \otimes U({\large g}) $ such that: 
\begin{equation}        
 C_{j}^{r}(\phi ,  \psi)= H_{j}^{r}(\phi \otimes  \psi).
\end{equation} \\
If we introduce the two homogeneous elements of degree zero 
of $U({\large g}) \otimes U({\large g})\lbrack \lbrack h \rbrack \rbrack $
$$F= 1+ \sum_{i \ge 1} F_{i} h^{i}$$
$$H= 1+ \sum_{j \ge 1} H_{j} h^{j}$$
then we obtain the following result
\begin{Proposition}
The associativity of the left-invariant  super star-product implies 
\begin{equation}        
 ( \Delta_{0} \otimes id)F . (F \otimes 1)=(1 \otimes  \Delta_{0} )F .(1 \otimes F)
\end{equation} 
and the associativity of the right-invariant super star-product leads to the following equality 
\begin{equation}        
 (S_{0}^{\otimes 2}(H) \otimes 1).( \Delta_{0} \otimes id)S_{0}^{\otimes 2}(H)  =(1 \otimes S_{0}^{\otimes 2}(H)).(1 \otimes \Delta_{0}  )S_{0}^{\otimes 2}(H) .
\end{equation}
\end{Proposition}
{\bf Proof}: writting the right-invariant super star product in the following form
$$ (\phi *^{r} \psi )= m (H^{r}(\phi \otimes \psi))$$
where $H= 1+ \frac{\rm h}{\rm 2} r + \sum_{i \ge 2} H_{i} h^{i} $\\

  we have for any homogeneous element $X$ in the enveloping superalgebra,\\ 

$< X, \phi *^{r} ( \psi *^{r} \chi)>$
$$ = <X, m(id \otimes m)((id \otimes \Delta_{0})H^{r}. H_{23}^{r}(\phi \otimes  \psi \otimes \chi))>$$
$$=<(id \otimes \Delta_{0})\Delta_{0}(X),(id \otimes \Delta_{0})H^{r}. H^{r}_{23}(\phi \otimes  \psi \otimes \chi)>   $$
\begin{equation} =<(1 \otimes (S_{0}^{\otimes 2}))H (id \otimes \Delta_{0})((S_{0}^{\otimes 2})H) (id \otimes \Delta_{0}) \Delta_{0} (X),(\phi \otimes  \psi \otimes \chi)>.\end{equation}

Similarly, we have

$< X,( \phi *^{r}  \psi) *^{r} \chi >$
\begin{equation} =<( (S_{0}^{\otimes 2}) \otimes 1 )H (\Delta_{0} \otimes id )((S_{0}^{\otimes 2})H) (\Delta_{0} \otimes id) \Delta_{0} (X),(\phi \otimes  \psi \otimes \chi)> \end{equation}

so, from (34) (35) we deduce easily the  result
(33).

 An analogous  proof establish the left-invariant case.\\ 
\begin{Proposition}
Assume that $F$ is a left-invariant super star product on the supergroup $G$, then $S_{0}^{\otimes 2}(F)$ is a right-invariant super star product on the supergroup $G$.
\end{Proposition}
{\bf Proof}: by applying the operator $(S_{0} \otimes S_{0} \otimes S_{0}  )$ to the equation (32) and using the fact that $(S_{0} \otimes S_{0}) \circ \Delta_{0}^{op} = \Delta_{0} \circ S_{0}$, we find obviously the equation(33).\\

 The  super star product on the  Poisson Lie supergroup will be given by the following expression
\begin{equation}        
 \phi * \psi= \mu((S_{0}^{\otimes 2})^{-1}(F^{-1})^{r}.F^{l}(\phi \otimes \psi))
\end{equation}
where $\mu$ is the usual mutiplication on the superalgebra of smooth functions on the supergroup. In fact,
the product defined in this way is associative

$$(\phi * \psi)* \chi =\mu((S_{0}^{\otimes 2})^{-1}(F^{-1})^{r}.F^{l}(\mu((S_{0}^{\otimes 2})^{-1}(F^{-1})^{r}.F^{l}(\phi \otimes \psi)) \otimes \chi))  $$
$$ =\mu(\mu \otimes id)(( \Delta_{0} \otimes 1)((S_{0}^{\otimes 2})^{-1}(F^{-1})^{r}).( \Delta_{0} \otimes 1)F^{l}.((S_{0}^{\otimes 2})^{-1}(F^{-1})^{r} \otimes 1).(F^{l}\otimes 1)(\phi \otimes \psi \otimes \chi))  $$
$$ =\mu(\mu \otimes id)(( \Delta_{0} \otimes id)((S_{0}^{\otimes 2})^{-1}(F^{-1})^{r}).((S_{0}^{\otimes 2})^{-1}(F^{-1})^{r}\otimes 1).( \Delta_{0} \otimes id)F^{l}.(F^{l} \otimes 1)(\phi \otimes \psi \otimes \chi))  $$
$$ =\mu(\mu \otimes id)(( id \otimes \Delta_{0} )((S_{0}^{\otimes 2})^{-1}(F^{-1})^{r}).(1 \otimes (S_{0}^{\otimes 2})^{-1})(F^{-1})^{r}.( id \otimes \Delta)F^{l}.(1 \otimes F^{l})(\phi \otimes \psi \otimes \chi))  $$
$$ =\mu( id \otimes \mu)(( id \otimes \Delta_{0} )((S_{0}^{\otimes 2})^{-1}(F^{-1})^{r}).(1 \otimes (S_{0}^{\otimes 2})^{-1})(F^{-1})^{r}.( id \otimes \Delta_{0})F^{l}.(1 \otimes F^{l})(\phi \otimes \psi \otimes \chi))  $$
$$ =\mu( id \otimes \mu)(( id \otimes \Delta_{0} )((S_{0}^{\otimes 2})^{-1}(F^{-1})^{r}).( id \otimes \Delta_{0})F^{l}.(1 \otimes (S_{0}^{\otimes 2})^{-1}(F^{-1})^{r}).(1 \otimes F^{l})(\phi \otimes \psi \otimes \chi))  $$ 
$ =\mu((S_{0}^{\otimes 2})^{-1}(F^{-1})^{r}.F^{l}.(\phi \otimes  \mu((S_{0}^{\otimes 2})^{-1}(F^{-1})^{r}. F^{l}(\psi \otimes \chi)))  $\\
$=\phi * (\psi* \chi ). $\\

For the compatibility relation, the proof is a graded version of the proof given in \cite{mo1}.

Actually a  super star-product does not only define a deformation of the superalgebra of the
 super smooth functions
on the supergroup ${\bf F}(G)$ but also of a quotient superalgebra ${\bf F}_{e}(G)$ 
 defined as the set of element of ${\bf F}(G)$  in a neighbourd containing the identity of $G$ modulo the  equivalence relation\\ 
$$ \phi \sim \psi ~~\mbox{ if}~~  < X, \phi - \psi > = 0  \mbox { for any}  X \in  U({\large g}), $$
where $ < , > $ is the pairing between ${\bf F}_{e}(G)$ and $ U({\large g}). $\\

Let us recall now that two bialgebras $U$, $A$ are said to be in duality if there
exists a doubly nondegenerate bilinear form 
	$$ < , >: U \times A  \longrightarrow  C ,  < , >: (u , a)  \longrightarrow  < u, a > ,  u \in U ,   a \in  A $$
such that for any $ u, v \in U$  and $ a, b \in  A  $ wa have:

	$$ < u ,ab >=  < \Delta_{A}(u) , a \otimes b > $$
	$$ < uv ,a >=  < u \otimes v , \Delta_{U}(a) > $$
	$$ < 1_{U} ,a >= \epsilon_{A}(a),   < u  ,1_{U} > = 
			 \epsilon_{U}(u)$$

All this extends to bisuperalgebras \cite{mani1}. The only subtlety is that the tensor product is also graded, and, if (using Sweedlers notation) $ \Delta_{U}(u) = \sum u_{1} \otimes u_{2}, \Delta_{A}(a) = \sum a_{1} \otimes a_{2} $, then 
$$ < u ,ab >=    (-1)^{ \mid u_{2} \mid  \mid a \mid } \sum < u_{1} ,a > < u_{2} ,b > $$  
$$ < uv ,a >=   (-1)^{ \mid a_{1} \mid  \mid v \mid } \sum < u ,a_{1} > < v ,a_{2} >$$
The duality between bisuperalgebras may be used to obtain the unknown superalgebra from
 a known one if the two are in duality.
So, the deformation we talk about is a deformation of the ${\bf F}_{e}(G)$ as a bialgebra
 ; this allows us to provide by the duality the deformed superalgebra ${\bf F}_{e}^{*}(G) \lbrack \lbrack h \rbrack \rbrack $ where ${\bf F}_{e}^{*}(G)$ is the set of distributions on $G$ with support at the the unit element $(e)$ . 
Indeed,  as in the case of ordinary Lie groups, the set of distributions on G with support
 at the identity element is the  enveloping superalgebra of the Lie  superalgebra of the
  Lie  supergroup, and we deduce that a  super star product provide a deformation 
of the enveloping  superalgebra.\\
The super quantized enveloping algebra $ U({\large g})\lbrack \lbrack h \rbrack \rbrack $ is endowed with the 
structure of a Hopf superalgebra where the multiplication superalgebra is the ordinary convolution on ${\bf F}_{e}^{*}(G)$  and  the coproduct $\Delta _{F}$ is given by \cite {ma1}
\begin{equation}        
 < \Delta _{F}(X), \phi \otimes \psi > = <X , \phi * \psi >  
\end{equation}
for all $\phi , \psi  \,\,\, \in \,\,\, {\bf F}_{e}(G)$, and $X \in U({\large g})$\\

 In fact using the equations (22),(23) we obtain:  
$$< \Delta _{F}(X), \phi \otimes \psi > = <X ,  \mu((S_{0}^{\otimes 2})^{-1}(F^{-1})^{r}.F^{l}(\phi \otimes \psi))>  $$ 
$$=  < \Delta _{0}(X) ,  (S_{0}^{\otimes 2})^{-1}(F^{-1})^{r}.F^{l}(\phi \otimes \psi)>  =  < F^{-1}. \Delta _{0}(X). F ,  (\phi \otimes \psi)>, $$
and this implies
\begin{equation}        
 \Delta _{F}(X) = F^{-1}. \Delta _{0}(X). F .
\end{equation}

For the antipode of the super quantized enveloping algebra, we recall first that the antipode $ S_{0} $ of $U({\large g})$ satisfies the following equation 

\begin{equation}        
m(S_{0} \otimes id ) \Delta_{0}(X)= m(id \otimes S_{0} ) \Delta_{0}(X)= \varepsilon (X)1
\end{equation} 
where $ m $ is the usual multiplication on the super enveloping algebra $U({\large g})$.\\
 $F$ and $F^{-1}$ can be respectively split as 
$$F = \sum _{k} a_{k} \otimes b_{k}, ~~~~  F^{-1}=  \sum _{k} c_{k} \otimes  d_{k} $$
and if seting  $u=  m(id \otimes S_{0} )(F^{-1})$ as an invertible homogeneous  element of $ U({\large g})\lbrack \lbrack h \rbrack \rbrack $ of degree zero, then we can easily show that the antipode of the super quantized enveloping algebra $U({\large g})\lbrack \lbrack h \rbrack \rbrack $ is given by:
\begin{equation}        
S_{F}(X)= u .S_{0}(X).u^{-1}
\end{equation} 
where $ u^{-1}= m(S_{0} \otimes id )F. $ \\
We will give the proof for the simple case when
$$ \mid a_{k} \mid = \mid b_{k} \mid = \mid c_{k} \mid = \mid d_{k} \mid = 0 $$
 since for others cases the generalization is obvious.\\
In fact.\\
$m(S_{F} \otimes id ) \Delta_{F}(X)
= m(u S_{0} u^{-1} \otimes id ) (F^{-1} \Delta_{0}(X) F)$
$$= \sum_{i,j,k} u S_{0}(a_{i})S_{0}(X'_{k})S_{0}(c_{j}) u^{-1} d_{j} X''_{k} b_{i}$$
with $ \Delta_{0}(X)= \sum_{k} X'_{k} \otimes X''_{k} $\\
Owing to the fact that $ S_{0}$ satisfies the equation(32) and that 
$$  \sum_{j} S_{0}(c_{j}) u^{-1} d_{j} =m(S_{0} \otimes id)(F. F^{-1}) = 1 $$
we obtain that
$$m(S_{F} \otimes id ) \Delta_{F}(X) =\sum_{i} u  S_{0}(a_{i})b_{i} \epsilon (X) 1= \epsilon (X) 1. $$

Similarly, we can prove that: 
$$ m( id \otimes S_{F} ) \Delta_{F}(X)= \epsilon (X) 1 $$

Now if we define the following  even element, as Drinfeld does in 
\cite{dr2} for the non graded case  
\begin{equation}        
 R_{F}= F_{21}^{-1}. F
\end{equation}
where $F_{21}= T.F_{12}.T$,  
then we can easily show that $R_{F} $ defines a quasitriangular structure on the super quantized enveloping algebra $U({\large g})\lbrack \lbrack h \rbrack \rbrack $.\\
In fact, applying the operator $ T^{23} T^{13}$ to the equation (32) and using the fact that $ \Delta^{op}_{0} = T \circ \Delta_{0}$ we obtain the following relation
$$F^{-1}_{12}(\Delta _{0} \otimes id )R_{F} F^{12}= (R_{F})_{13}.(R_{F})_{23}$$
which implies that 
\begin{equation}        
 (\Delta _{F} \otimes id )R_{F}= (R_{F})_{13}.(R_{F})_{23}.
\end{equation}
Similarly, applying $T^{12} T^{23} $ to the same equation(32),we obtain
\begin{equation}        
 ( id \otimes \Delta _{F})R_{F}= (R_{F})_{13}.(R_{F})_{12}.
\end{equation}
From the fact that 
$ \phi * 1 = 1* \phi = \phi $ for all $ \phi \,\, \in {\bf F}_{e}(G)$, we deduce that 
\begin{equation}        
 ( id \otimes \varepsilon)F = ( \varepsilon \otimes id)F =1;
\end{equation}
consequently 
\begin{equation}        
 ( \varepsilon \otimes id)(R_{F}) =( id \otimes \varepsilon)(R_{F}) = 1
\end{equation}
and from the definition(41) we deduce that 
\begin{equation}        
 (R_{F})_{21}.R_{F} = 1.
\end{equation}
Using  the expression (38) we obtain that :
$$ ( \Delta _{F})^{op} = T( \Delta _{F}) = T(F^{-1}).\Delta _{0}.T(F)$$
$$ =T(F^{-1}).F.\Delta _{0}.F^{-1}.T(F)$$ 
then 
\begin{equation}
(\Delta _{F})^{op}=R_{F}.\Delta _{F}.(R_{F})^{-1}.
\end{equation}
From (43) and (47), we see that $R_{F}$ satisfies the super quantum Yang-Baxter equation
\begin{equation}
(R_{F})_{12}.(R_{F})_{13}.(R_{F})_{23}=(R_{F})_{23}.
(R_{F})_{13}.(R_{F})_{12}.
\end{equation}

\subsection{ Representation of  the Super Quantum Yang Baxter Equation}

\hspace*{0.5cm}  Consider a graded space $ W^{(n/m)}  $ consisting of n bosons and m fermions. Let $ \rho $ be a  representation of the  Lie superalgebra $ {\large g} $ on  $W^{(n/m)}$, then 
$$ R =( \rho \otimes \rho )(R_{F}) \,\,\, \in
(W^{(n/m)} \otimes W^{(n/m)}) $$
satisfies the  super quantum Yang-Baxter equation
\begin{equation}        
R_{12}.R_{13}.R_{23}= R_{23}. R_{13}.R_{12}.
\end{equation}
If we choose $ \{ w_{i} \}$ as a basis of $W^{(n/m)}$ , where\\
$ \mid w_{i} \mid = 0$  for $ i= 1,2,.....,n$ \\
$ \mid w_{i} \mid = 1$  for $ i= n+1,n+2,.....,n+m $ \\
then the  equation(49)can be rerwitten as  :

$$
 (-1)^{\mid m \mid  (\mid c \mid + \mid n \mid )  } (-1)^{\mid d \mid  (\mid m \mid + \mid n \mid + \mid e \mid + \mid f \mid  )} R_{i m}^{ab}.R_{dn}^{ic}.R_{e f}^{mn} $$

$$
= (-1)^{\mid m \mid  (\mid k \mid + \mid f \mid ) } (-1)^{\mid a \mid  (\mid b \mid +\mid c \mid + \mid m \mid + \mid k \mid  )} R_{m k}^{bc}.R_{lf}^{ak}.R_{de}^{lm}$$
where $ \mid i \mid = \mid w_{i} \mid $, and if we introduce
the matrix  $S = P R $ , where $P$ is the super permutation operator on the tensor vector space
$ W^{(n/m)} \otimes  W^{(n/m)}$ with
$$ P^{ij}_{kl}= (-1)^{ \mid i \mid  \mid j \mid } \delta^{i}_{l}  \delta^{j}_{k} $$ 
then $ S $ satisfies  
$$      
 (-1)^{\mid a \mid  ( \mid b \mid + \mid c \mid \mid j \mid +\mid n \mid )}
(-1)^{\mid d \mid  (\mid m \mid + \mid m \mid + \mid n \mid + \mid e \mid +  \mid f \mid )} S_{jm}^{bc}.S_{dm}^{aj}.S_{ef}^{mn} $$
\begin{equation}
= (-1)^{\mid l \mid ( \mid j \mid + \mid c \mid  \mid m \mid + \mid f \mid   )}  S_{l j}^{ab}.S_{mf}^{jc}.S_{de}^{lm} 
\end{equation}
which can be rewritten in a compact form as 
\begin{equation}        
(S \otimes id).(id \otimes S).(S \otimes id)= (id \otimes S). (S \otimes id).(id \otimes S).
\end{equation}
This gives rise to a representation of the symmetric group $ S_{n}$

\subsection{Equivalents Super  Star Products on a Supergroup}
\hspace*{0.5cm}  Let $F$ and $ \bar F $ be two  super star-products i.e, two homogeneous elements of degree zero of the  Hopf superalgebra
$(U({\large g})\lbrack \lbrack h \rbrack \rbrack )$ and let 
$A = U(({\large g})\lbrack \lbrack h \rbrack \rbrack), \Delta_{F}, R_{F}, S_{F}  )$ and $ \bar A =( U(({\large g})\lbrack \lbrack h \rbrack \rbrack), \Delta_{\bar F}, R_{\bar F}, S_{\bar F} )$

be the resulting  quantum supergroups, where 
$$\Delta_{F}= F. \Delta_{0}.F^{-1} ,~~~~
 R_{F}= F_{21}^{-1}. F $$
$$\Delta_{\bar F}= \bar F. \Delta_{0}.\bar F^{-1} ,~~~~
R_{\bar F}=\bar F_{21}^{-1}. \bar F $$ 
then it is easily seen that $ \bar A $ can be obtained from A by applying the twist $ \hat F  = F^{-1}. \bar F $. In fact 
\begin{equation}        
\Delta_{\bar F} = \hat F. \Delta_{F}. \hat F^{-1}
\end{equation}
and
\begin{equation}        
R_{\bar F} = \hat F_{21}. R_{F}. \hat F .
\end{equation}
If the two  star product are equivalent i.e. the corresponding elements 
$F$ and $ \bar F $are related by the following expression
\begin{equation}        
\bar F =\Delta_{0}(E^{-1}). F. (E \otimes E)
\end{equation}
for some invertible homogeneous element E of degree zero  of  
$U({\large g})\lbrack \lbrack h \rbrack \rbrack $, then the coproduct $ \Delta_{\bar F} $ can be rewritten as
\begin{equation}        
\Delta_{\bar F}(X) = (E^{-1} \otimes E^{-1}) \Delta_{F}(E.X.E^{-1}).(E \otimes E)
\end{equation}
Similarly, the quasitringular structures are related by
\begin{equation}        
R_{\bar F} = (E^{-1} \otimes E^{-1}). R_{F}.(E \otimes E)
\end{equation} 
And  the two twisted antipodes are related by the following expression 
\begin{equation}        
S_{\bar F} = E^{-1} S_{0}(E^{-1}). S_{F}.S_{0}(E).E.
\end{equation} 

Then the inner automorphism of degree zero of the superalgebra
structure $ E(.)E^{-1}$ defines now a Hopf superalgebra isomorphism of degree 
zero. Finally, from (56) we see that the induced isomorphim of degree zero 
maps the quasitriangular structures into each other as well.\\

\end{document}